\documentclass[conference]{IEEEtran}
\IEEEoverridecommandlockouts

\usepackage{amsmath}

\usepackage{graphicx}
\usepackage{csquotes}
\usepackage{algorithm2e}
\usepackage{paralist}
\usepackage[table,xcdraw]{xcolor}
\usepackage{pdfpages}

\usepackage[style=ieee]{biblatex}
\usepackage[section,numberedsection=autolabel]{glossaries}
\makeglossaries

\newglossaryentry{function}{name=function,%
    description={A single, self-contained piece of software, that performs a certain function}}

\newglossaryentry{feature}{name=feature,
    description={Composed of one or more functions connected together using a certain runtime environment, usually corresponds to a certain use-case}}

\newglossaryentry{runtime environment}{name=runtime environment,
    description={Communication middleware and virtualization mechanisms}}

\newglossaryentry{ECU}{name=ECU,
    description={Electronic Control Unit. It is an electronic device in a vehicle that is responsible for a single function}}

\newglossaryentry{TDD}{name=TDD,
    description={Test-Driven Development is a software development methodology that centers on the iterative creation of unit tests prior to the implementation of functional code~\cite{ref40:Beck2002} This approach mandates that a test case specifying the desired behavior of a code unit be written before the production code itself. As development progresses, the test suite continuously executes. New code is only written if it fulfills the requirements outlined in a failing test}}

\newglossaryentry{FDD}{name=FDD,
    description={Feature-Driven Development. It is a paradigm where the software system is iteratively developed in a series of steps, starting with an abstract model of the system, followed by extraction of a set of desired features, and ending with feature implementation and integration~\cite{ref41:Palmer2001}}}

\newglossaryentry{MBSE}{name=MBSE,
    description={Model-Based Systems Engineering is a formalized methodology within systems engineering that emphasizes using models as the primary means of information exchange and system representation~\cite{ref42:Incose2023}. This contrasts with traditional document-centric approaches. MBSE centers on creating and leveraging domain-specific models or metamodels, which capture system requirements, design, analysis, and verification elements throughout the development lifecycle}}

\newglossaryentry{contract}{name=contract,
    description={Design by contract is a software development methodology that emphasizes the explicit definition of formal contracts between software components~\cite{ref43:Mitchell2002}. These contracts specify preconditions (what must be true before a component is used), postconditions (what must be true after execution), and invariants (conditions that must always hold true). Design by contract can be enforced through runtime assertions, unit tests, or even integrated into a programming language's syntax. This approach enhances software reliability, eases debugging, and facilitates code comprehension}}

\newglossaryentry{metamodel}{name=Metamodel,
    description={Defines the language of system description by specifying abstract entities that are part of the system, a set of possible relations between them, and their attributes}}

\newglossaryentry{instance model}{name=Instance model,
    description={A model generated from the given metamodel, populated with actual objects with concrete attribute values; an implementation of the system described in the language of the metamodel}}

\newglossaryentry{OMG}{name=OMG,
    description={Object Management Group}}

\newglossaryentry{LLM}{name=LLM,
    description={Large Language Model}}

\newglossaryentry{OCL}{name=OCL,
    description={Object Constraint Language}}

\newglossaryentry{RACE}{name=RACE,
    description={Centralized Platform Computer Based Architecture for Automotive Applications}}

\newglossaryentry{Ecore}{name=Ecore,
    description={Language of the metamodel used in Eclipse Modeling Framework}}

\newglossaryentry{OEM}{name=OEM,
    description={Original Equipment Manufacturer}}

\makeatletter % changes the catcode of @ to 11
\newcommand{\linebreakand}{%
  \end{@IEEEauthorhalign}
  \hfill\mbox{}\par
  \mbox{}\hfill\begin{@IEEEauthorhalign}
}
\makeatother % changes the catcode of @ back to 12

\title{Synergy of Large Language Model and Model Driven Engineering for Automated Development of Centralized Vehicular Systems \\\vspace*{20pt} \normalsize  \today{}}

\author{
\IEEEauthorblockN{Nenad Petrovic\IEEEauthorrefmark{1}}
\IEEEauthorblockA{email: nenad.petrovic@tum.de \\
orcid: 0000-0003-2264-7369}
\and
\IEEEauthorblockN{Fengjunjie Pan\IEEEauthorrefmark{1}}
\IEEEauthorblockA{email: f.pan@tum.de \\
orcid: 0009-0005-8303-1156}
\and
\IEEEauthorblockN{Krzysztof Lebioda\IEEEauthorrefmark{1}}
\IEEEauthorblockA{email: krzysztof.lebioda@tum.de \\
orcid: 0000-0002-7905-8103}
\and
\IEEEauthorblockN{Vahid Zolfaghari\IEEEauthorrefmark{1}}
\IEEEauthorblockA{email: v.zolfaghari@tum.de \\
orcid: 0009-0004-0039-6014}
\and
\IEEEauthorblockN{Sven Kirchner\IEEEauthorrefmark{1}}
\IEEEauthorblockA{email: sven.kirchner@tum.de \\
orcid: 0009-0004-3845-6772}
\and
\IEEEauthorblockN{Nils Purschke\IEEEauthorrefmark{1}}
\IEEEauthorblockA{email: nils.purschke@tum.de \\
orcid: 0009-0008-9470-5795}
\and
\IEEEauthorblockN{Muhammad Aqib Khan\IEEEauthorrefmark{1}}
\IEEEauthorblockA{email: aqib.khan@tum.de \\
orcid: 0000-0002-1082-253X}
\and
\IEEEauthorblockN{Viktor Vorobev\IEEEauthorrefmark{1}}
\IEEEauthorblockA{email: vorobev@in.tum.de \\
orcid: 0000-0002-0473-148X}
\and
\IEEEauthorblockN{Alois Knoll\IEEEauthorrefmark{1}}
\IEEEauthorblockA{orcid: 0000-0003-4840-076X}
\linebreakand
\IEEEauthorrefmark{1}\IEEEauthorblockA{Technical University of Munich (TUM) \\
School of Computation, Information and Technology (CIT) \\
Chair of Robotics, Artificial Intelligence and Embedded Systems}
}

\date{\today}

\begin{document}
\null%
\includepdf{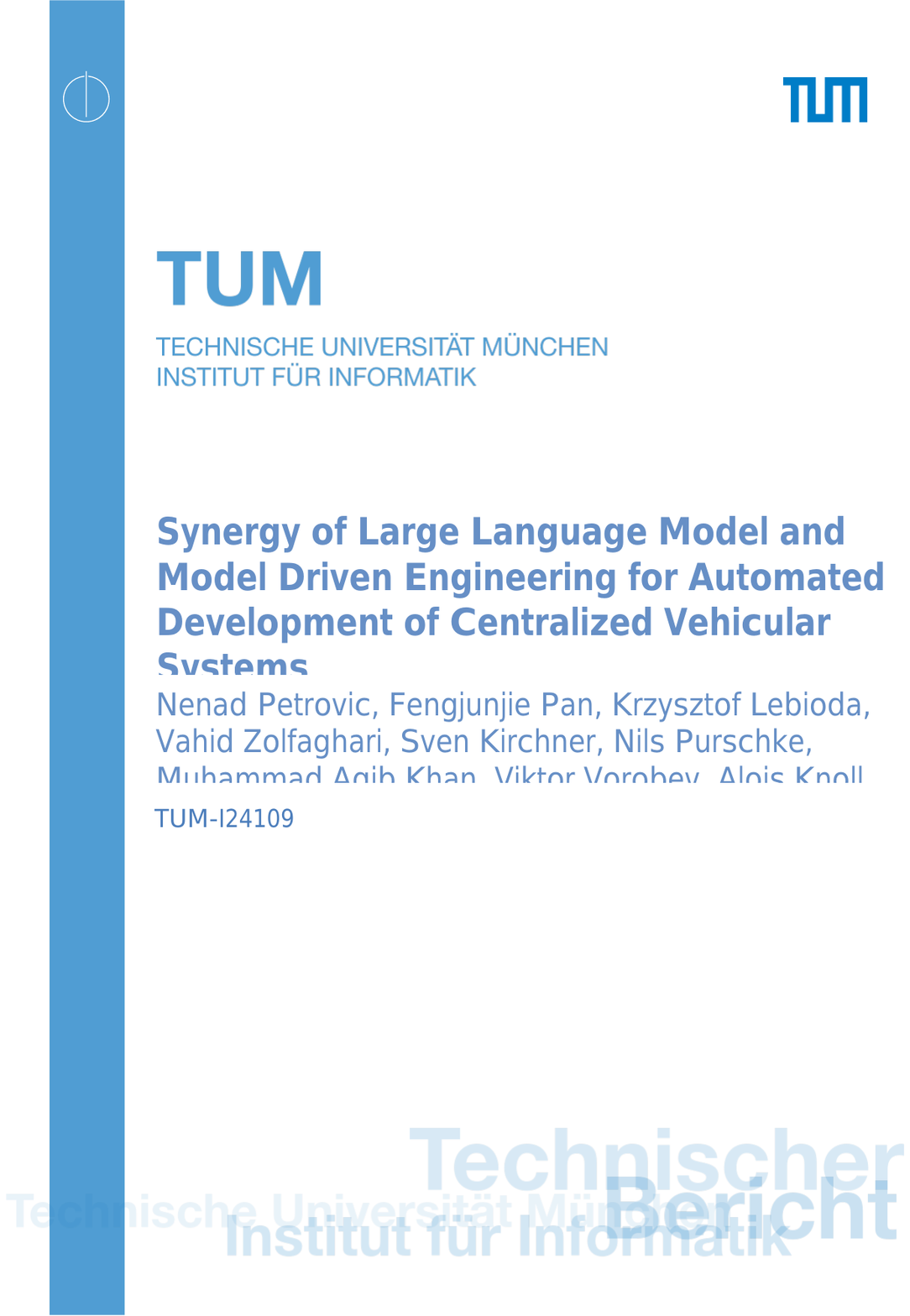}

\maketitle

\begin{abstract}
We present a prototype of a tool leveraging the synergy of model driven engineering (MDE) and Large Language Models (LLM) for the purpose of software development process automation in the automotive industry. In this approach, the user-provided input is free form textual requirements, which are first translated to Ecore model instance representation using an LLM, which is afterwards checked for consistency using Object Constraint Language (OCL) rules. After successful consistency check, the model instance is fed as input to another LLM for the purpose of code generation. The generated code is evaluated in a simulated environment using CARLA simulator connected to an example centralized vehicle architecture, in an emergency brake scenario.
\end{abstract}

\section{Introduction}

The rise of large language models (LLMs), initiated by popularity of the now well-known ChatGPT towards the end of 2022, has affected almost every aspect of our everyday life -- from simple customer services and question answering to entertainment and science. Apart from its innovation, the main drive behind LLM adoption in various areas working field and scientific areas is to either automatize repetitive tasks which are traditionally done by humans, or help them achieve improved outcomes by providing suggestions or hints. In area of computer science and software engineering the usage of LLMs is identified for several distinct purposes in literature \cite{ref45:Petrovic2023, ref46:Petrovic_AlAzzoni_2023}: 
\begin{compactenum}
    \item \textbf{code generation} -- automatically generating of either boilerplate code or full applications, algorithms and other useful code scripts (tests, configuration files);
    \item \textbf{code augmentation} -- adding the missing parts to the code or additional functions/classes to achieve the given goal;
    \item \textbf{code analysis} -- providing suggestions how the provided source code can be improved in order to avoid well-known bugs, security vulnerabilities and other code-related issues which are identifiable in design-time;
    \item \textbf{code explanation} -- LLM is able to provide quite verbose comments which correspond to specific parts of the code, explaining its purpose and usage;
    \item \textbf{data analysis} -- it was also shown that LLMs trained on enormous amount of textual data are also able to mimic the traditional machine learning algorithms and perform supervised learning tasks, such as classification.
\end{compactenum}

On the other side, when it comes to automotive industry, software is the crucial aspect which shapes the innovation in this area. However, traditional automotive architectures rely on hundreds of heterogeneous devices (electronic control units -- ECUs), provided by different OEMs, using distinct architectures, OS and software platforms. Therefore, software development and its maintenance in automotive area becomes increasingly difficult, taking into account the demand for additional functionalities (such as infotainment) and advanced automated control scenarios (such as autonomous driving), which usually involve the usage of even more devices, such as sensors, actuators and specific co-processors.
The latest concept of centralized automotive architecture enables a flexible software-defined vehicle.

Therefore, in this paper, we propose the means for LLM-aided design and developments of centralized car systems, starting from high-level feature specification. Apart from LLMs, our proposed framework relies on model-driven software engineering principles. Utilization of formal specifications -- such as metamodels and constraint rules -- enables verification and validation of LLM outcomes, which are, due to their nature, prone to the so-called hallucinations, where the generated output is incorrect. Such unreliability can be fatal in domains such as automotive. Therefore, we propose design-time, formal verification of generated results. In synergy with user supervision, and manual intervention and correction where needed, it should largely mitigate the cons of LLM usage. 

In this work, LLMs will be adopted to tackle the following aspects:
\begin{compactenum}
    \item requirements extraction and summarization -- generating formal list of verifiable requirements based on free-form ISO standard or reference architecture text;
    \item automated design -- creating verifiable car system model instances based on user requirements;
    \item code generation -- automated parametrization of deployment configuration templates.
\end{compactenum}

\section{Motivation behind centralized architecture}

%\begin{figure}[ht]
 %   \includegraphics[width=\linewidth]{workflow.pdf}
  %  \caption{Proposed workflow with generative AI in the loop. Steps are color-coded: 1) Inputs in orange, outputs in red - generation of the instance model and formal constraints. The formal constraints are used for automatic verification of the instance models; 2) Inputs in red, outputs in blue - resource allocation, which produces an allocation matrix and enhances the instance model with details about software-to-hardware mapping; 3) Inputs in blue, outputs in purple - code generation. The tests generated during this step are used for automatic testing of the deployed system; 4) Yellow - validation and verification, which is performed as part of the other steps, and is not a separate step as such.}
   % \label{fig1}
%\end{figure}

The main goal of a centralized car architecture is to provide the so-called "single-system illusion", which has many advantages compared to other state-of-art solutions \cite{ref17:Sommer2013, ref47:Tas_2017}.

\textbf{Lower Hardware Costs}: Centralized systems consolidate multiple control functions into powerful processors, significantly reducing the need for numerous individual control units with separate processors, memory, and electronics. This consolidation can lead to economies of scale in purchasing, resulting in lower overall hardware costs. Additionally, the amount of computing power can be minimized through efficient scheduling of the different functionalities on a centralized system. In contrast, in distributed systems redundant computing hardware exists since there is a need to process data on different hardware components that are often not shared across the system, especially if components from multiple manufacturers are involved. This leads to parts of the hardware being idle instead of working on other processing tasks.

\textbf{Improved Energy Efficiency}: By minimizing the number of control units and optimizing computing resources, centralized architectures can decrease vehicle's energy consumption. This efficiency is crucial for electric vehicles, where energy usage directly affects range and performance. Since there is only the central unit that requires power for data processing, it is much easier to monitor and influence it.

\textbf{High Application-Level Communication Speed}: Centralization allows for faster communication speeds at the application level, as data does not have to traverse multiple nodes (control units) before reaching its destination. Instead, the applications run on the same computational hardware and can directly access any data without the need to communicate with external components.

\textbf{Simplicity in Application Development}: Writing applications becomes simpler in a centralized system, because developers only need to target a single computing environment, streamlining the development process. All applications can be unified in terms of inter-application-communication, sensor and actuator abstraction through the use of an underlying hardware abstraction layer, isolation mechanisms like containers, scheduling systems and so on. This vastly simplifies the software development and makes it largely hardware independent.

\textbf{Simplified Software Failure detection}: Since all software runs on a single system, it is much easier to identify occurring errors. All systems can be monitored on a software level instead of the need to debug the communication on the wire. What is more, the amount of proprietary, black box ECUs from various manufacturers is minimized. Uniform runtime environment simplifies error handling since common low level error detection routines can be used, which monitor all applications and handle errors on the same platform using unified error handling mechanisms, without the external communication overhead.

\textbf{Full System Control}: A centralized architecture ensures that all vehicle data and processing capabilities are centralized, providing comprehensive control over the vehicle's functionality through software. The amount of private intellectual property, including custom hardware and software, in the vehicle is reduced. Therefore, many software components and functionality can be adapted by the car manufacturer itself, without the need of help from external suppliers.

\textbf{Simplified Wiring and Connectivity}: Direct connections to the central computational unit or straightforward data tunneling through network switches eliminate the complex wiring and communication layouts, characteristic to distributed systems.

\textbf{Facilitates Software-Defined Vehicle Concepts}: Centralizing data and processing simplifies the transition towards software-defined vehicles, where software, rather than hardware, defines vehicle functionality. This centralization supports easy software updates and upgrades, feature improvements, and addressing safety issues.

\textbf{Ease of Hardware Upgrades}: If designed with hardware extensibility in mind, centralized systems simplify the addition of new hardware, contrasting with the challenges of integrating new components in hardware-heavy distributed systems, where the hardware is often hardwired with the rest of the system and permanently installed on the chassis, meaning that any hardware update or upgrade is a complex task. Additionally, when exchanging components of a distributed system, the compatibility with the rest of the system must be ensured, which may be difficult if the components were heavily interconnected. A centralized hardware built on top of proper hardware abstraction mechanisms is largely unchallenged by those problems.

\section{LLM-Enabled Workflow}
\begin{figure*}[t]
    \centering
    \includegraphics[width=0.8\textwidth]{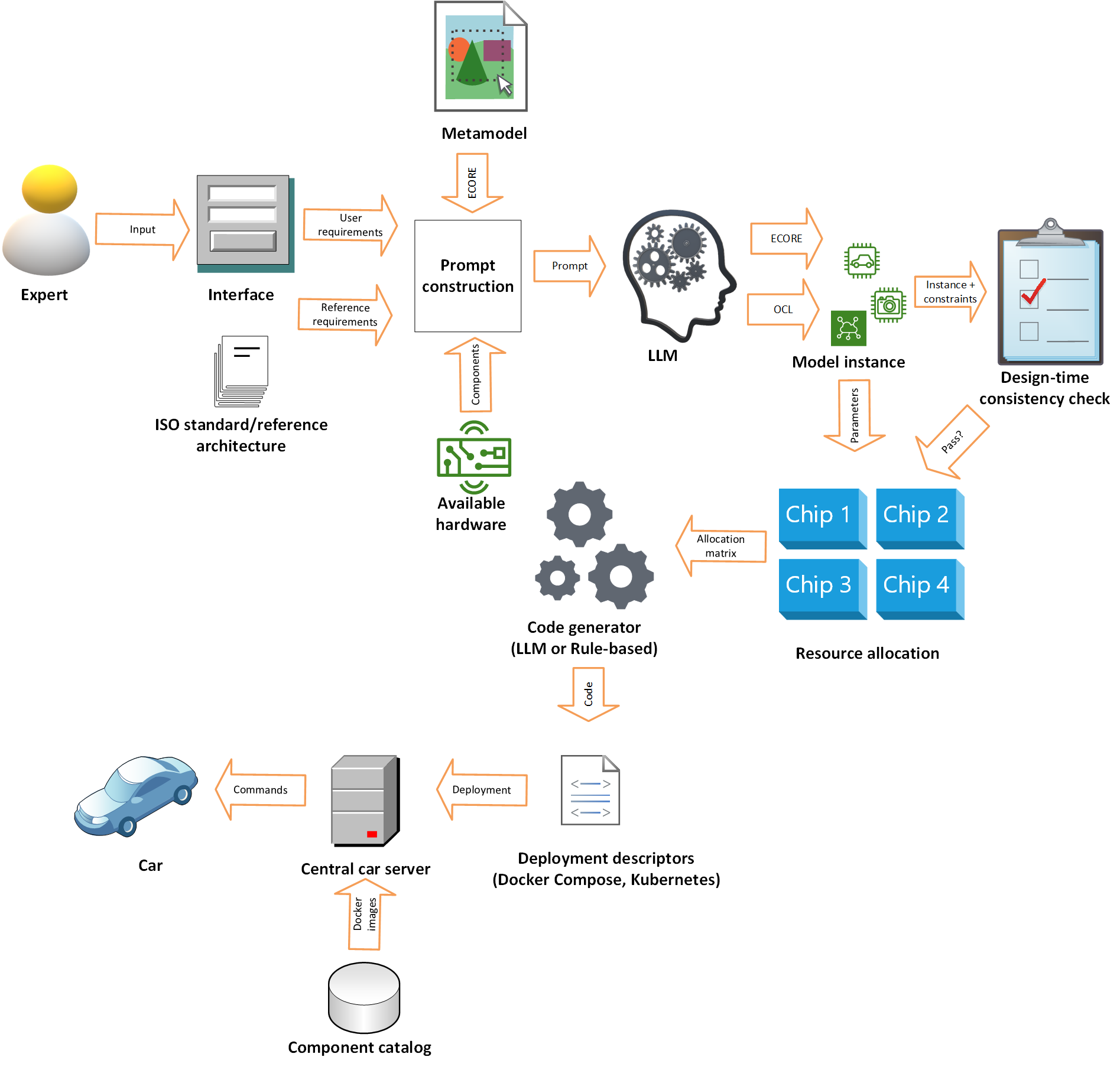}
    \caption{LLM-enabled workflow for model-driven automotive development.}
   \label{fig1}
\end{figure*}

The proposed workflow leveraging synergy of LLMs and model-driven engineering is depicted in Fig. \ref{fig1}.

In the first step, a domain expert enters the requirements in free-form natural language into the given form with pre-defined fields. The arguments from this form are forwarded, together with the predefined Ecore metamodel, to a prompt constructor (described later). The prompt for the LLM is constructed using the following template:

\textit{Prompt1 -- Model instance creation: Create XMI model instance for [Requirements input] and [Ecore metamodel]}

As outcome of the process, an XMI model instance depicting the requirements with respect to the Ecore metamodel is generated.

For the purpose of requirements consistency verification, OCL rules \cite{ref30} based on a reference architecture and the metamodel are extracted as well:

\textit{Prompt2 -- Constraints extraction: Based on reference requirements [Reference requirements or part of ISO standard document] and metamodel [Ecore metamodel] extract OCL constraints}

Once the instance model is verified, resource allocation step is preformed, which assigns car features to corresponding chips available to the central car server. The outcome of this process is an allocation matrix giving the information on how features and their corresponding containers will be mapped to the hardware.

The created model instance is leveraged for code generation, either using conventional rule-based solutions or an LLM. There are several possible steps where model instance can be used for code generation: parametrizing pre-defined components (CARLA excerpts \cite{ref52}), containerizing components, generating deployment descriptors. The corresponding parameters are extracted from model instance and inserted into prompt templates, or the whole metamodel/model instance is used.

\begin{figure*}[t]
    \centering
    \includegraphics[width=0.8\textwidth]{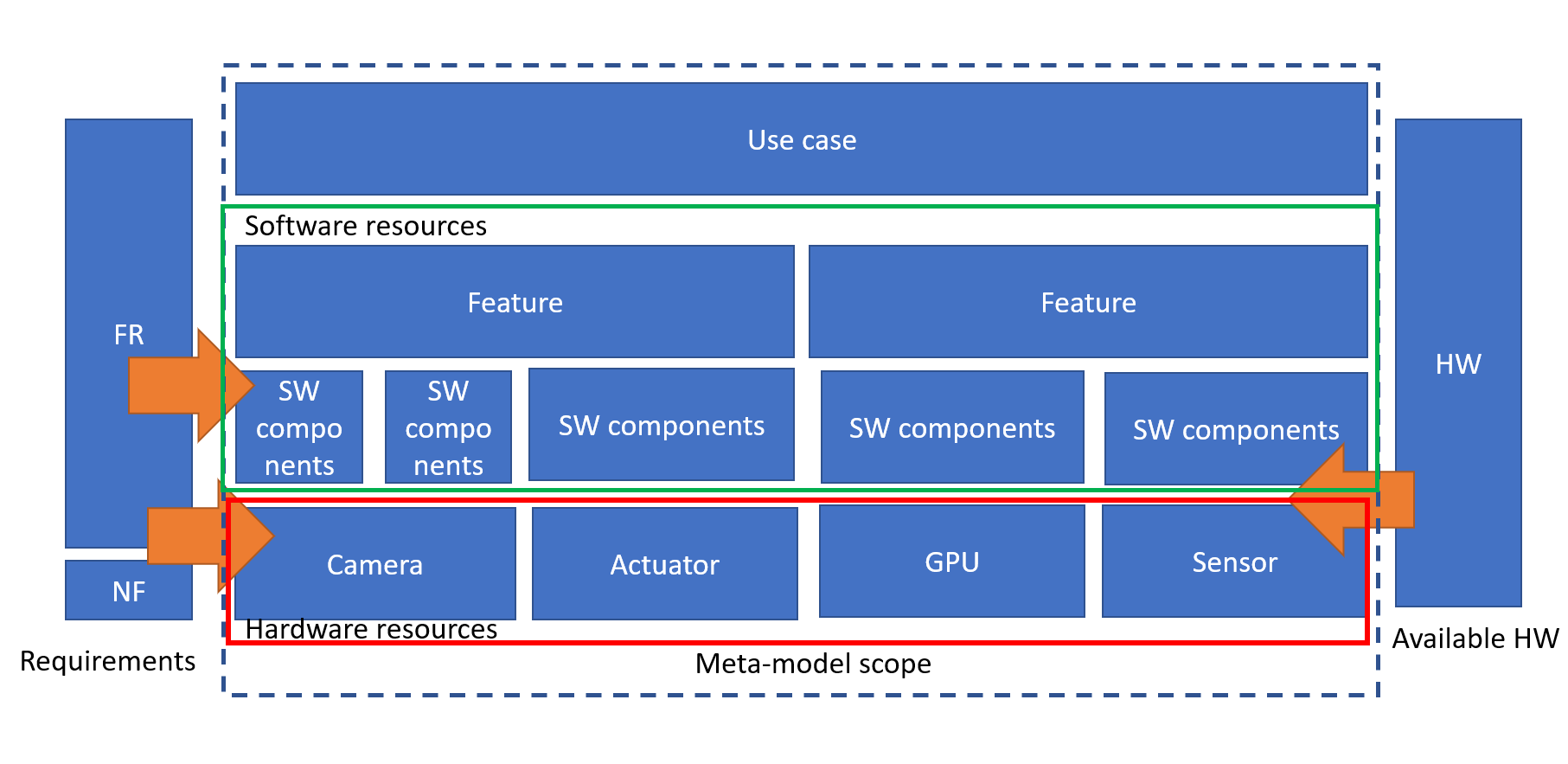}
    \caption{Hierarhical representation of Centralized Car Server Metamodel.}
   \label{fig2}
\end{figure*}

\textit{Prompt3 -- Parametrizing components: Based on Ecore instance [Ecore instance model]
fill in the template [CARLA excerpt]}

The goal of this prompt is to leverage parameters from Ecore model instance, describing car component characteristics and add insert them CARLA code excerpt.

\textit{Prompt4 -- Containerizing components: Write Dockerfile to containerize script [script name] written in [Language/Technology used] using dependencies [Libraries and so on]}

\textit{Prompt5 -- Automated generation of adapters between components: Generate Python code to convert [source format] to [target format]}

\textit{Prompt6--Deployment descriptor generation: Generate [Docker Compose or Kubernetes descriptor] to run containers:
[For each container in ZoneController.containers:]
[Container.image] on {ZoneController} open port [container.communicationPort] with environment variables [Container-\textgreater Sensor/Acutator properties] based on model instance [Ecore model instance]}

Finally, the generated Docker Compose file or Kubernetes descriptor is deployed and run on Central Car Server. The corresponding Docker images (Python scripts relying on CARLA client library in our case) are assumed to be available within local or custom (remote) Docker image repository. Customization of these docker images is performed by parametrizing the corresponding Docker Compose environment variables, based on sensor/actuator characteristics. 

The whole code generation workflow from user requirements to CARLA code is described in form of pseudocode in \ref{alg:one}.

\begin{algorithm}
\caption{Code generation}\label{alg:one}
\KwData{$requirement$, $metamodel$, $standard$}
\KwResult{$CARLAcode$}
$xmi \gets executePrompt(prompt1, requirement, metamodel)$\;
$ocl \gets executePrompt(prompt2, standard, metamodel)$\;
$pass \gets verify(ocl, xmi, metamodel)$\;

{\eIf{$pass$ is true}{
$CARLAcode \gets executePrompt(prompt6, xmi)$\;}
{$CARLAcode \gets null$\;
$print(Requirement not compliant with standard)$}
}
\end{algorithm}

Additional details and whole vision of full LLM integration, including user feedback steps are described in \cite{ref51:Lebioda}. However, they are not included in the presented prototype implementation, but planned to be included in the final version of the tool.

\section{Centralized Car Server Metamodel}

Centralized Car Server metamodel covers both the hardware and software components, together with aspects of functional (FR) and non-functional (NF) requirements. Hierarchical overview of aspects covered by metamodel is given in Fig. \ref{fig2}.

As it can be seen, on the top level, metamodel covers the use cases/scenarios in automotive as a concept, such as highway pilot. These scenarios are further broken down into features, which are mapped to the actual software components, represented as Docker containers in our case. On the other side, when it comes to resource allocation, there are two different perspectives involved: 1) Software component allocation - how features are mapped to containers 2) Hardware component allocation - how software components are mapped to set of available hardware resources. For the first aspect, the crucial elements are functional requirements which determine the set of features that will be taken into account. On the other side, when it comes to aspects of hardware allocation, non-functional requirements, such as costs, power consumption, processing power and others are considered.

This metamodel is based on the idea that Central Car Server-controlled car contains several \textit{ZoneControllers}, where each of them is responsible for distinct part of car (like front, rear, side etc). However, from the developer's perspective, the architecture is designed to provide illusion of unified, centralized system. ZoneController could be either an isolated virtual machine running within Central Car Server (therefore, it is important to take into account which VM platform and image are used) or physically distinct application-specific chip - Co-Processor (such as Infineon's solution).

\textit{ProcessingNode} represents element of metamodel that generalizes components which are able to execute processing tasks. For such elements, the following aspects are relevant and taken into account:

\begin{itemize}
    \item  memory capacity
    \item processing power
    \item number of cores
    \item maximum bandwidth/data rate
    \item architecture - ARM/x86 etc.
    \item realtimeCapability - whether the processing node has support for realtime execution
\end{itemize}

Within the system's scope, we identify two types of processing node components: Master and Slave. While slave components are usually single, dependent chips (required to be connected to a master) - such as GPU, TPU and FPGA, master components represent more general boards with additional components which can act independently, attach/detach slave devices and schedule tasks for execution. In that sense, ZoneControllers are subclass of master processing nodes.

To each ZoneController, a set of Components (either Hardware or Software) is assigned in order to implement particular usage Scenario. Moreover, scenario consists of one or many Features. Each of the Features is derived from free-form textual specification coming from Requirements Specification. Moreover, a feature can be safety-critical, which is reflected in container redundancy and other strategies adopted to ensure safety compliance by the components.
When it comes to hardware components there are three identified types: Co-Processors, Sensors and Actuators. Co-Processor represents slave ProcessingNodes, such as GPU, TPU, FPGA which are designed to execute some specific tasks with high performance, such as object detection and sensor data analysis. Sensors cover commonly used sensing devices in vehicular systems, such as Camera, Lidar, Ultra-violet and others. Role of these devices is to record quantitative measurements representing some real-phenomenon.

For \textit{Sensors}, the following aspects are relevant:
\begin{itemize}
    \item measurementUnit - such as pixels, intensity
    \item measurementFormat - RGB bitmap, CSV data and other
    \item measurementsPerSecond- how frequently the data is measured
    \item parameterList - list of sensor-specific parameters, such as Camera's fov angle, image Height and Width, LIDAR's number of channels etc.
\end{itemize}

On the other side, for \textit{Actuator}, relevant aspects are:
\begin{itemize}
    \item commandFormat - such as MQTT JSON message or some other specific protocol
    \item parameters - list of command parameters, such as acceleration amount, steering angle
\end{itemize}

Additionally, to each of the Sensors/Actuators, corresponding software container Controller is assigned in runtime, which is run within the isolated environment on the ZoneController. This container is responsible for enabling the function of sensor/actuator device. For sensors, it can start sensing, performing reading sensor measurements and stop sensing. For actuators, it can turn the device on or off and issue commands to them.

When it comes to \textit{Software} resources, they have the following common properties:
\begin{itemize}
    \item memoryDemand - estimation how much memory is required on the assigned ProcessingNode to run it
    \item processingDemand - estimated processing power required
    \item bandwidthDemand - how much data is exchanged or consumed by the software component
    \item realtimeRequired - denotes whether the software components require realtime capabilities
\end{itemize}

The main unit of software abstraction is \textit{ApplicationContainer}. It represents an isolated environment where some specific processing task or sensor/actuator controller is executed. Container has the following properties:
\begin{itemize}
    \item image - name of the image which is used as base for creation of container
    \item targetTechnology - which kind of underlying containerization/orchestration engine is used, such as Docker, Docker Compose or Kubernetes engine
    \item repository - URL of the corresponding online repository, private container registry or automotive vendor software catalogs  where resources required for running the container can be accessed
    \item script [optional] - exact name of the script which will be run inside the container, useful only for additional capabilities of automated script containerization
    \item dependencies [optional] - list of libraries and runtime environments (such as Python) used by container, also useful only in case of automated containerization
    \item architecture - for which processor architecture was the container built for, such as ARM/x86
    \item communicationPorts - list of ports which are used for communication with this container, so they have to be exposed in order to access them from other components, such as port 2000 in case of CARLA simulator-based components
\end{itemize}

Additionally, one more type of Software components are \textit{ProcessingTasks}. They represent either data analysis method implementaton (image or LIDAR-measurement based object detection) or control algorithms. ProcessingTasks usually take data as input from either Sensor element or another ProcessingTask. For each of them, the following aspects need to be taken into account:
\begin{itemize}
    \item inputs - list of aspects which are taken as inputs of the performed computation/processing/transforamtion. For example, it is "image" for object detection.
    \item outputs - what is generated as outcome of the performed transformation, such as steering or brake command in case of car control algorithm
    \item inputFormat - list of formats corresponding to each of the inputs specified, such as "RGB" or "CMYK" for image
    \item outputFormat - list of formats for the products generated
    \item compatibleFormat - list of compatible formats, recognized by task as proper input. In case that provided inputs are not in corresponding format, adoption of adapter methods between the components would be needed   
\end{itemize}

Regarding the abstract interface of these components, they perform processing which takes into account inputs, inputs formats and output formats, while the result of the transformation outcome is output. There are various possibilities for input and output combinations:
\begin{itemize}
    \item Image based object detection - inputs are images, outputs is a number of objects detected
    \item Control algorithm - input is a number of obstacles, outputs are acceleration, steering angle or brake activation
    \item ConnectionLink elements are used for representation of relationships - either virtual or physical connections between two components, considering the protocol, type of connection and latency.
\end{itemize}

The aspects of power consumption, together with costs by each of the hardware components (co-processors, sensors and actuators) can be taken into account as well, especially in the design phase when the decision regarding the selection of corresponding parts have to be taken into account by non-functional requirements-aware optimization-based resource allocation mechanisms.

\begin{figure*}[ht]
    \centering
    \includegraphics[width=\textwidth]{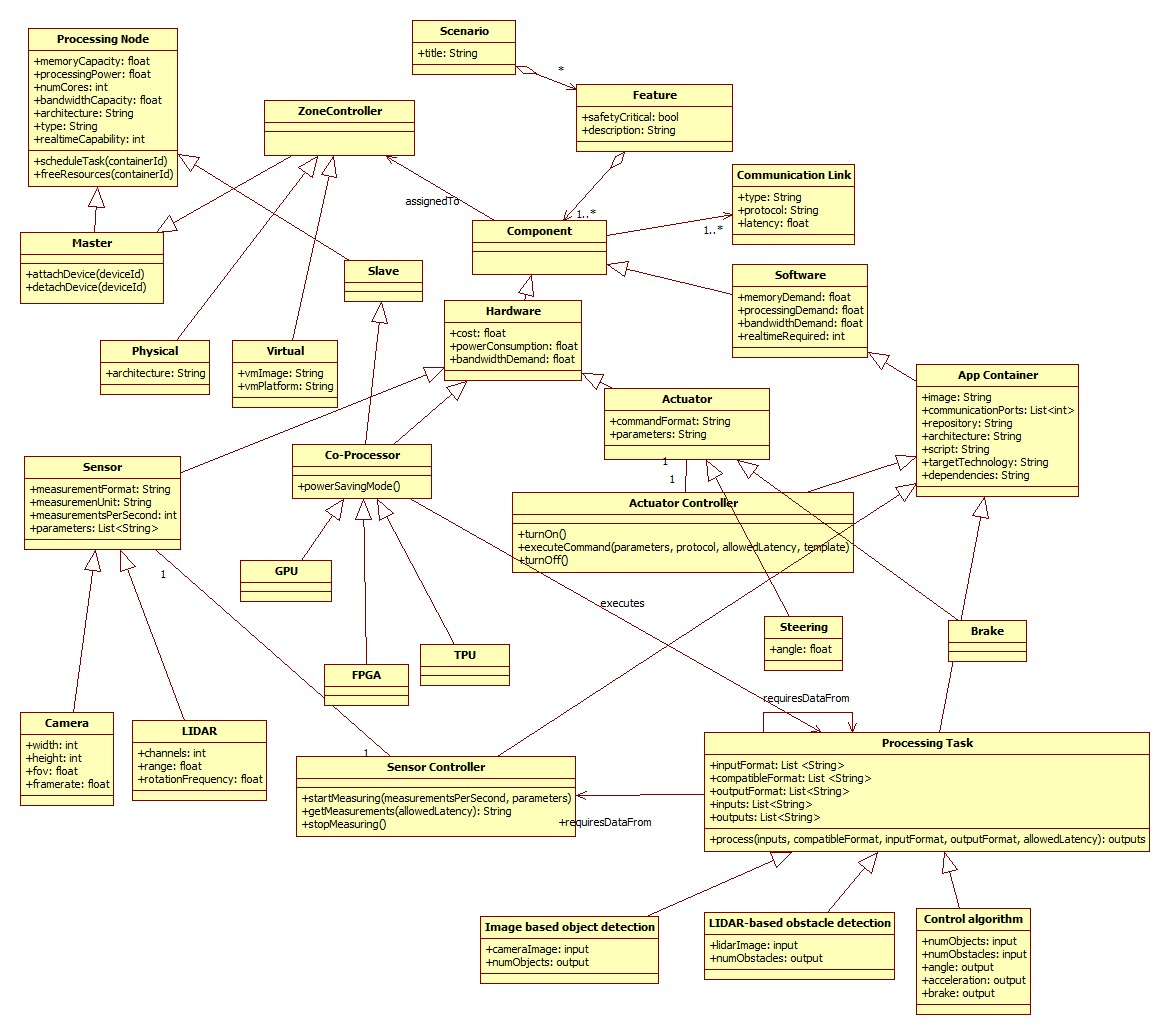}
    \caption{Centralized Car Server Metamodel.}
   \label{fig3}
\end{figure*}

\section{Component Interfaces}
This section gives an overview of high-level, abstract interfaces characteristic for various types of system component interfaces.

\texttt{ProcessingNode}
\begin{itemize}
    \item \texttt{scheduleTask(containerId)} -- processing nodes should be able to execute the assigned computational tasks
    \item \texttt{freeResources (containerId)} -- processing nodes should be able to free up the resources for app containers which are no longer required
\end{itemize}

\texttt{ZoneController} (inherited from \texttt{ProcessingNode $\rightarrow$ Master}): Apart from task scheduling, ZoneCntrollers should be able to attach/detach slave devices, such as sensors, actuators and co-processors
\begin{itemize}
    \item \texttt{attachDevice(deviceId)}
    \item \texttt{detachDevice(deviceId)}
\end{itemize}

\texttt{Actuator Controller}: application container responsible for managing of actuator devices (turn on or off) and execution of commands for given protocol, format and specific parameters.
\begin{itemize}
    \item \texttt{turnOn()}
    \item \texttt{turnOff()}
    \item \texttt{executeCommand(protocol, parameters, template, allowedLatency)} -- parametrizes command template with given set of actuator-specific parameters, while the deadline for command completion should not exceed the value of allowed latency
\end{itemize}

\texttt{Sensor Controller}: application container enabling the acquisition of measured sensor data and managing of sensing devices
\begin{itemize}
    \item \texttt{startMeasuring(measurementsPerSecond, parameters)} -- sets up sensing device to start measuring the observed real world phenomenon, generating the given number of measurements per second, for given sensor setup configuration parameters
    \item \texttt{stopMeasuring()} -- device should stop measuring values
    \item \texttt{getMeasurements(allowedLatency)} -- component should return the current measurements in no longer time than the one given by \texttt{allowedLatency} parameter
\end{itemize}

\texttt{Processing Task}: application container which takes sensor data or other processing task's output values as its input, performs some kind of data transformation and generates outputs in given format
\begin{itemize}
    \item \texttt{process(inputs, inputFormat, compatibleFormat, outputFormat, allowedLatency): outputs}
\end{itemize}

\texttt{Co-Processor}: additional chips used for processing should be able to go to power saving mode if not used for computation
\begin{itemize}
    \item \texttt{powerSavingMode()}
\end{itemize}

Interfaces are not included as part of Ecore metamodel specification. Therefore, another formalism is used for interface specification, such as Interface Definition Language (IDL). Example of adopting such notation for ZoneController is given:
\begin{verbatim}
interface ZoneController{
    void scheduleTask(String containerId);
    void freeResource(String containerId);
    void attachDevice(String deviceId);
    void detachDevice(String deviceId);
}
\end{verbatim}

\section{Prototype implementation}

The initial implementation of prompting tool is done in Python programming language. When it comes to construction of prompts that will be forwarded to LLM, we can distinguish four crucial processing steps:
\begin{itemize}
    \item model parsing -- depending on particular scenario involving LLM, different information, such as property values could be required from the model instances. For that purpose, considering that we rely on Ecore framework, the corresponding parsing library for Python is used -- PyEcore. This library enables convenient model instance parsing, element traversal and parameter retrieval.

    \item prompt construction -- in this step, we take the parsed values from the model instance or user-provided free-form text and parametrize pre-defined prompt templates.
    
    \item prompt execution -- invoking the prompting service API of the underlying LLM -- either locally deployed or in cloud (assuming that the corresponding LLM is properly deployed, accessible via current network and exposes such interface). As we make use of ChatGPT in our current experiments, OpenAI's Python API for this LLM-based service is used.

    \item interpreting LLM outputs -- post-processing of the results which are returned from LLM, including their parsing, extraction of relevant results and even further transforming them to suitable form for the next block within the LLM-enabled pipeline. In this step, it is of utmost importance to clear the LLM-generated result from all the additional text which resulted due to "hallucination" effect. Therefore, after this step, we should have clean, at least syntactically correct output of LLM which can be further used.
    \end{itemize}
    
\begin{figure*}[t]
    \centering
    \includegraphics[width=\textwidth]{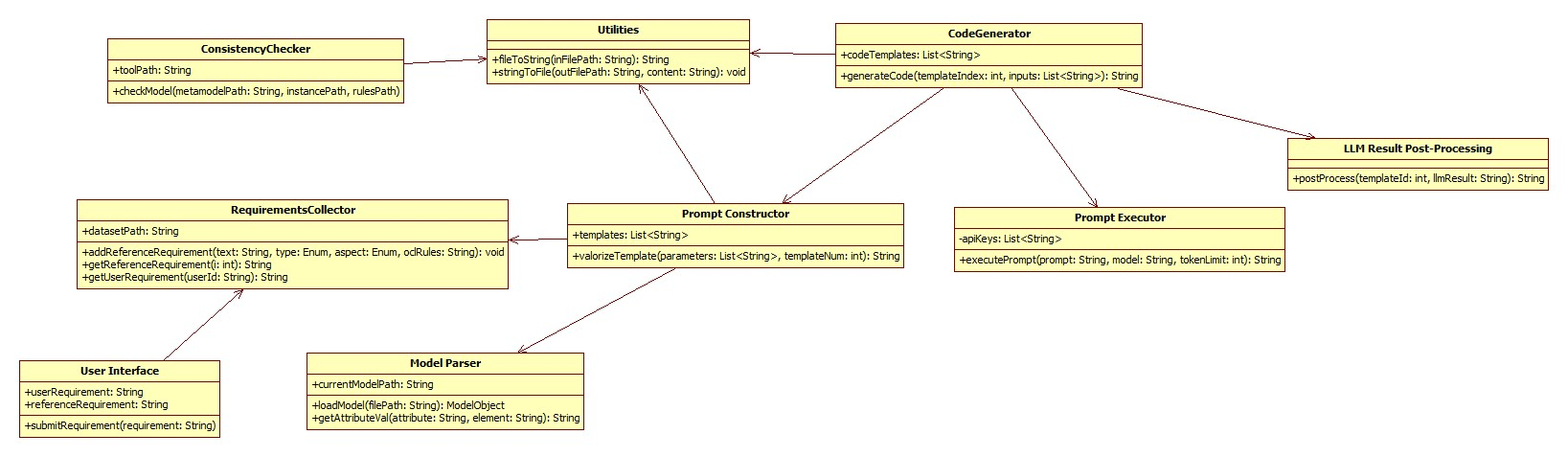}
    \caption{LLM-enabled tool prototype.}
   \label{fig4}
\end{figure*}

UML class diagram illustration the prototype tool implementation is given in Fig \ref{fig4}.

In the first step, Ecore model instance file is loaded into the tool and parsed using \texttt{loadModel} method of \texttt{ModelParser} which has \texttt{currentModelPath} as argument. As outcome of this process object representation is formed in memory, so its structure and properties can be traversed. After that, it is possible to retrieve the values of distinct element attributes, using \texttt{getAttribute} method, which looks for the given model instance element and returns the value of target attribute/property. Moreover, we have \texttt{PromptConstructor} which contains a list of prompt templates. In our current version of the prototype, we include three types of templates, which are relevant to the scope of our proposed workflow: model instance creation, OCL rule extraction and code generation template.

The implementation of the prompt construction mechanism is in \texttt{valorizeTemplate}, which parametrizes the prompt template from the list with given identifier value. The second argument of this method is the list of parameter values which will be used to parametrize the prompt selected by id. These parameters are extracted from the model instance itself, relying on \texttt{ModelParser}. 

Once the prompts are constructed, the next step is \texttt{PromptExecutor}, which is responsible for executing the prompts parametrized with appropriate values from the instance model. \texttt{PromptExecutor} can optionally have one or more \texttt{apiKeys} stored, which are used for authentication in case of external LLM-based services, such as ChatGPT. These values are user-specific and should not be exposed publicly. The crucial method of prompt executor is \texttt{executePrompt} which sends the constructed prompt to selected LLM model, taking into account the maximal answer length, limited by \texttt{tokenLimit} (which prevents from unexpected charges in case of external, pay per use services). The result of this method is raw output coming directly from LLM in response to the forwarded prompt.

\texttt{CodeGenerator} coordinates the LLM-based flow -- from user-defined and reference requirements in textual form to generated executable code, such as Docker container. It contains a list of code templates for various types of assets that are expected to be generated (such as Docker Compose YAML file, Kubernetes YAML descriptors or CARLA Python scripts) which are inserted as input to prompt constructor for parametrization of specific prompts.

However, the generated response could contain redundant or additional text which is not useful for the further workflow, so it needs to be post-processed. For this purpose, we perform post-processing of the generated results and the outcome depends on the template identifier itself, as the same template is likely to lead to the generation of similar result by LLM. The underlying method is \texttt{postProcess}, and it has two arguments: \texttt{templateId} -- identifier of prompt template; \texttt{llmResult} -- the raw output of LLM. The output of this method is the string which contains only the useful information, without additions.

For intermediary steps involving model instance usage, before the process of code generation, the XMI Ecore model instance is forwarded to \texttt{ConsistencyChecker}. Its main method performs XMI model instance verification for given Ecore metamodel definition and set of OCL rules. For each of the rules, the output would be pass or fail. In case that all checks pass, the flow goes further to code generation phase. Otherwise, a message to the end-users is sent, so they know that manual intervention or correction is needed.  In our prototype, this class relies on an external executable file implemented in Java programming language and packed as JAR file.

Additionally, our current prototype implementation also includes several auxiliary classes:
\begin{itemize}

    \item \texttt{UserInterface} -- Handles user input from web form - either reference or custom requirements, and forwards them to \texttt{RequirementsCollector}.

    \item \texttt{RequirementsCollector} -- Used for construction of requirements dataset in automated manner. It is assumed that web-based GUI is used by experts in order to enter the requirements in textual form with respect to the structure, which is later described in the section about the used datasets. There are three methods in this class. The first one adds the requirement to the dataset in CSV format, based on reference requirements. The second retrieves the i-th row from the requirements. The third one accepts user-defined requirement, coming from the web input form.

    \item \texttt{Utilities} -- Provides functionalities enabling to read textual file as string from disk and vice versa -- to write string variable into given textual file.
\end{itemize}

\section{Experiment demonstration environment}
The Central Car Server runs isolated environments for ZoneControllers, where each of the ZoneControllers runs a set of assigned Docker containers. ZoneControllers can be either physically (distinct chips or co-processors) or logically isolated (VMs or containers).

Practically, to each of the ZoneControllers, a distinct Docker Compose file defining the running containers and ports they use for communication can be assigned. Additionally, adoption of Kubernetes is considered as well, especially considering the cloud-over-the-air update scenario and redundancy mechanism incorporation in order to increase the overall robustness of the resulting system. When it comes to code generation, the idea is to extract model instance parameters and target to produce Docker Compose files or Kubernetes descriptor YAML files, where each of these files will be executed by one of the zone controllers.

\begin{figure}[ht]
    \includegraphics[width=\linewidth]{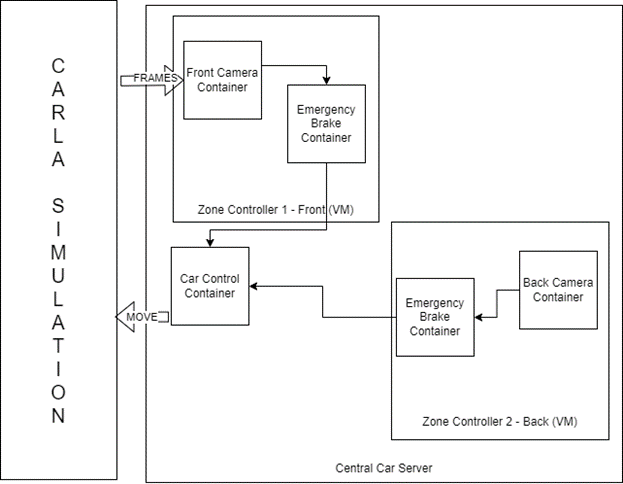}
    \caption{Integration of central car server with CARLA simulation.}
   \label{fig5}
\end{figure}

\begin{figure}[ht]
    \includegraphics[width=\linewidth]{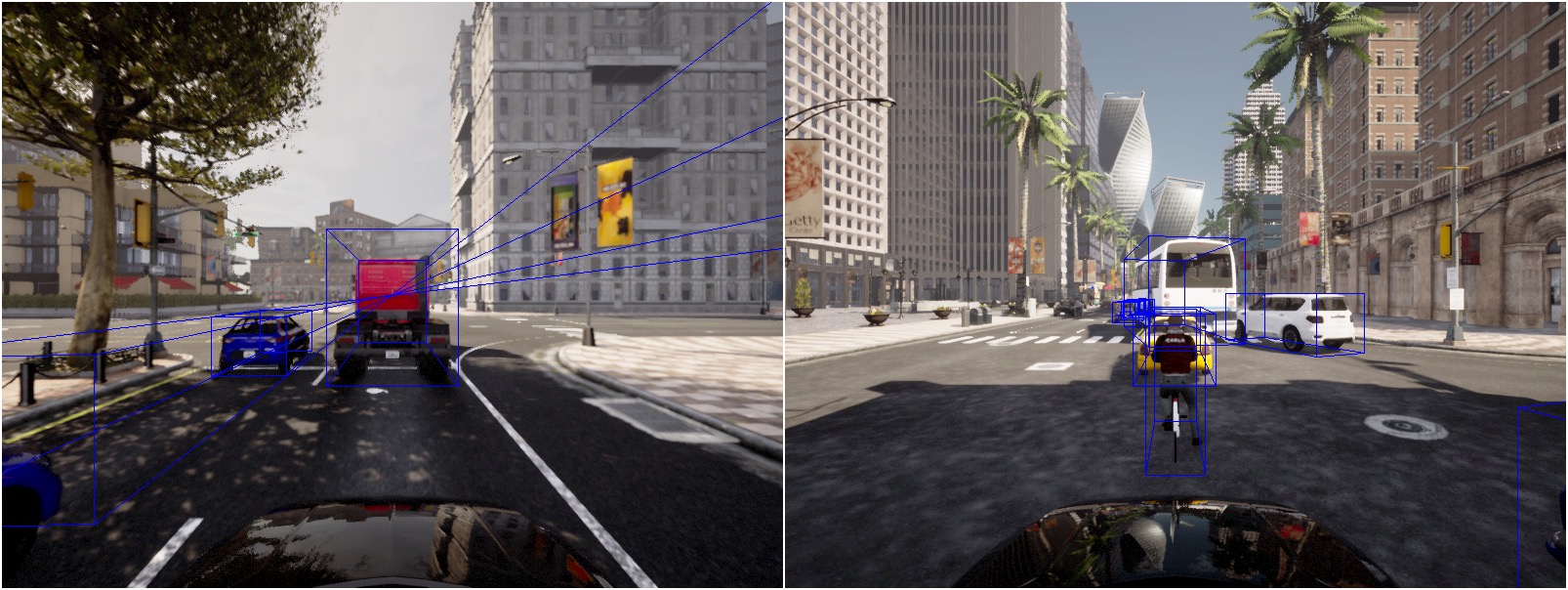}
    \caption{3D Object Detection in CARLA simulation environment for central car server evaluation.}
   \label{fig7}
\end{figure}

In the given example (shown in Fig. \ref{fig5}), we have two ZoneControllers – back and front. The first one is responsible for the frontal part of the car and runs camera sensing container which produces input for object detection-based emergency brake. In case that object is detected at short distance, automatic emergency brake will be activated and send signal to central car control container, as shown. A basic demonstration of object detection implemented in CARLA simulator using one virtual camera in the 3D environment is shown in Fig. \ref{fig7}. Similarly, there is a distinct Zone Controller for the back part of the vehicle.
In our demonstration, it is assumed that two-way integration with CARLA simulation is established, so the images are recorded from CARLA simulation environment, while commands issued by control component affect the car’s position in simulation. The communication port in our case is 2000 when it comes to communication with CARLA simulator server in both directions.

\begin{figure}[ht]
    \includegraphics[width=\linewidth]{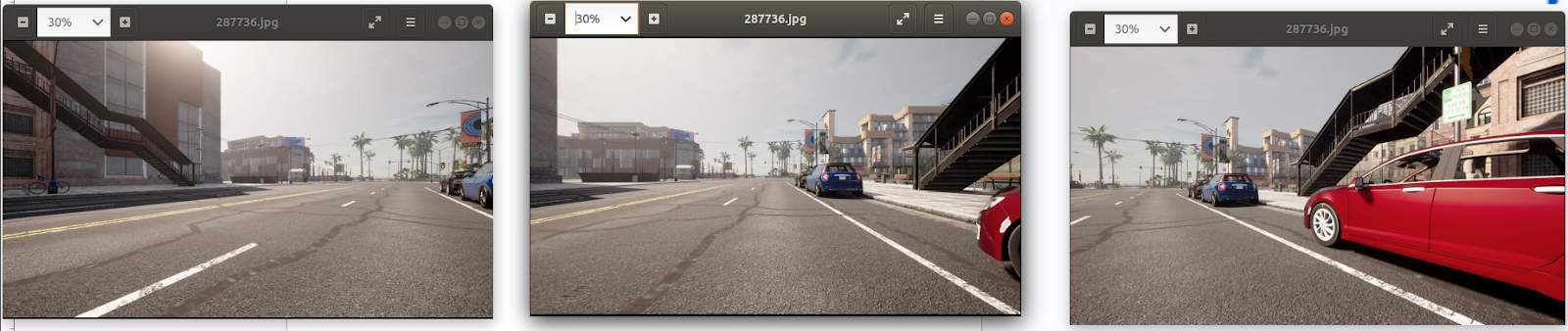}
    \caption{CARLA simulation environment for central car server evaluation.}
   \label{fig6}
\end{figure}

A screenshot of CARLA simulator, showing the perspective of multiple virtual cameras in 3D environment, is given in Fig. \ref{fig6}.

\section{Resource allocation}
The resource allocation problem in this case can be treated as a mapping of software resources: $m$ application containers ($c_1$, $c_2$, ..., $c_m$) to $n$ processing nodes ($p_1$, $p_2$, …, $p_n$) with respect to some pre-defined criterion policy, such as minimal price, minimal power consumption, minimal latency or maximum execution speed. 

In what follows, the developed model for optimal deployment is described, built upon the works from \cite{ref44:Pan2022} and \cite{ref34:Azzoni2021}. For each pair of processing node and container, we assign a decision variable $allocation[i,j]$ indicating whether the container $c_j$ is deployed on the node $p_i$ or no:

\begin{equation}
    allocation[i,j] =
    \left\{
        \begin{aligned}
            &1,\quad c_j~\textrm{is deployed on}~p_i, \\
            &0,\quad \textrm{otherwise.}
        \end{aligned}
    \right.
\end{equation}

An example objective function would be to minimize the cost of components required, by deploying the containers to the cheapest nodes that fulfill all the criteria:

\begin{equation}
    minimize \sum_{i=1}^{n} {\sum_{j=1}^{m} {cost[i]\cdot allocation[i,j] }},
\end{equation}
where $cost[i]$ is the price of the processing node $p_i$.

Furthermore, the following constraints have to be considered. First, the sum of the application container memory demands -- $memoryDemand[j]$ -- of the allocated containers should not exceed the memory capacity of the processing node -- $memoryCapacity[i]$, where this container is deployed, given as:

\begin{multline}
    \sum_{j=1}^{m} {allocation[i,j] \cdot memoryDemand[j] } \leq \\
    \leq memoryCapacity[i], \quad \forall i\in[1,n]
\end{multline}

Similar conditions hold for bandwidth constraints and processing power:
\begin{multline}
    \sum_{j=1}^{m} {allocation[i,j] \cdot bandwidthDemand[j] } \leq \\
    \leq bandwidthCapacity[i], \quad \forall i\in[1,n]
\end{multline}

\begin{multline}    
    \sum_{j=1}^{m} {allocation[i,j] \cdot processingDemand[j] } \leq \\
    \leq processingPower[i], \quad \forall i\in[1,n]
\end{multline}

Each container, which requires real-time capabilities, must be deployed on the processing node with such support:

\begin{multline}
    \sum_{j=1}^{m} allocation[i,j] \cdot \\
    \cdot \left|realtimeCapability[i] - realtimeRequired[j]\right| =0, \\ 
    \forall i\in[1,n]
\end{multline}

Similarly, for each container, its architecture type (x86, ARM, GPU, TPU) must match with the processing node’s computing architecture, which is relevant in our case as various co-processors can be adopted as well

\begin{multline}
    \sum_{j=1}^{m} allocation[i,j] \cdot \\
    \cdot \left|nodeArchitecture[i] - appArchitecture[j]\right| =0, \\
    \forall i\in[1,n]
\end{multline}

\section{Experiments}
This section gives overview of the examples relying on LLM usage, including distinction between the inputs, output and the underlying prompt for each of them.
\subsection{Model instance creation}

\begin{figure*}[t]
    \centering
    \includegraphics[width=\textwidth]{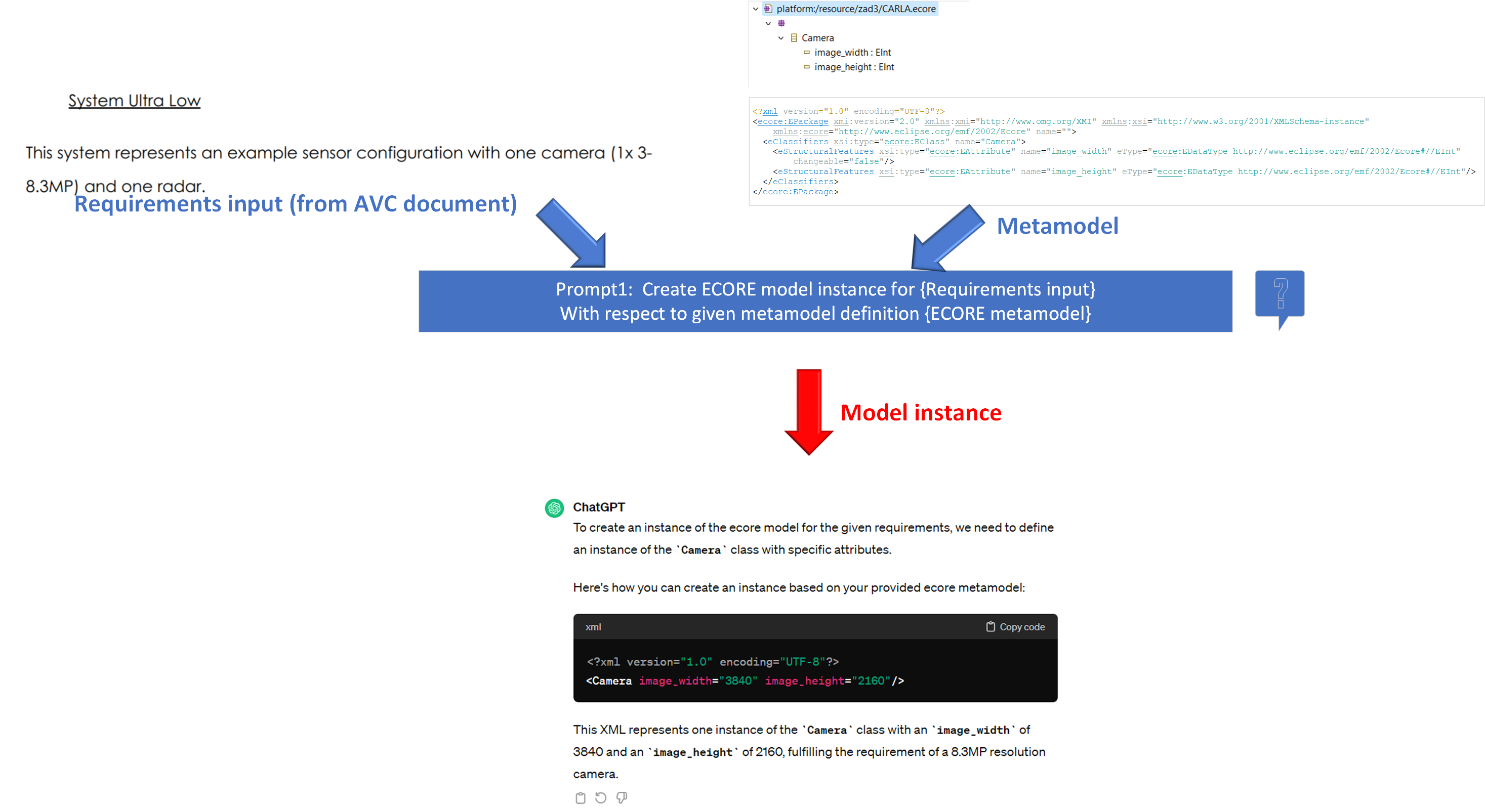}
    \caption{Model instance creation experiment illustration.}
   \label{fig8}
\end{figure*}

Requirements + Metamodel $\rightarrow$ Model instance (Fig. \ref{fig8})

%\begin{figure}[ht]
%    \includegraphics[width=\linewidth]{exp2.png}
%    \caption{Model instance creation experiment illustration.}
%   \label{fig8}
%\end{figure}

\subsection{Constraint extraction}
\begin{figure*}[t]
    \centering
    \includegraphics[width=\textwidth]{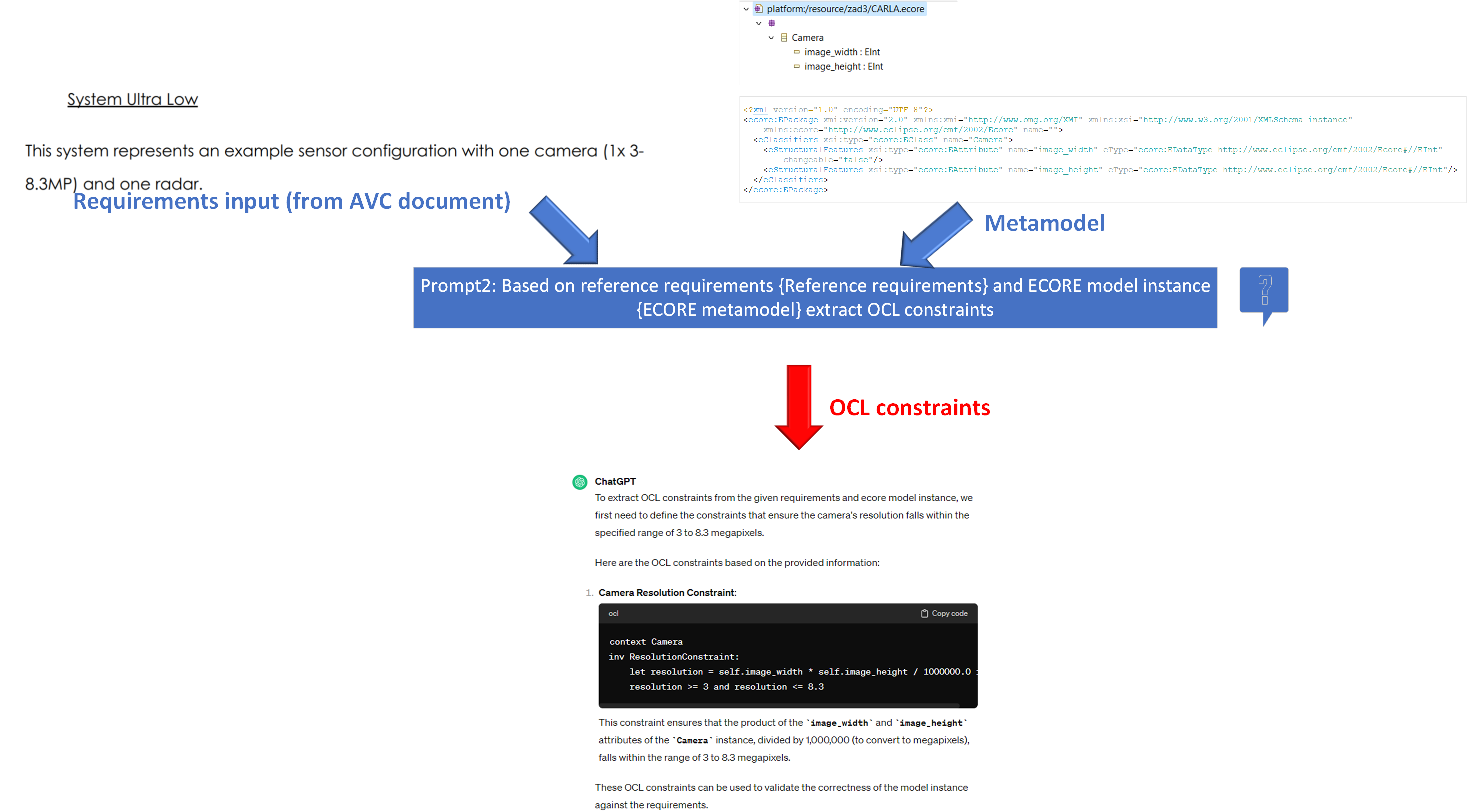}
    \caption{OCL constraint rule extraction experiment illustration.}
   \label{fig9}
\end{figure*}

Requirements + Metamodel $\rightarrow$ OCL constraints (Fig. \ref{fig9})

%\begin{figure}[ht]
%    \includegraphics[width=\linewidth]{exp5.png}
%    \caption{OCL contraint rule extraction experiment illustration.}
%   \label{fig8}
%\end{figure}

\subsection{CARLA code generation}
\begin{figure*}[t]
    \centering
    \includegraphics[width=0.8\textwidth]{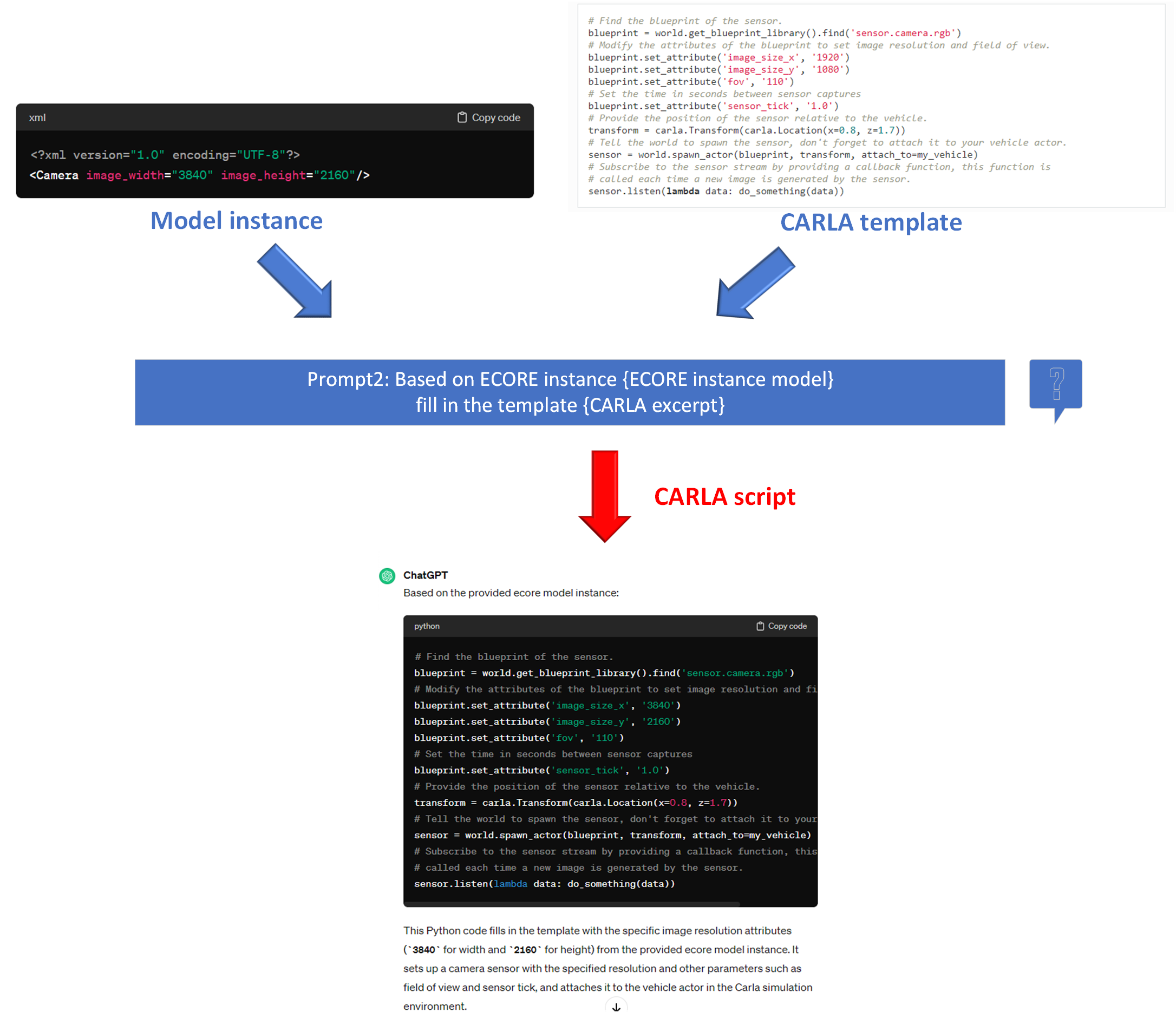}
    \caption{CARLA code generation experiment illustration.}
   \label{fig10}
\end{figure*}

Model instance + Template $\rightarrow$ Component Script (Fig. \ref{fig10})
%\begin{figure}[ht]
%    \includegraphics[width=\linewidth]{exp4.png}
%    \caption{CARLA code generation experiment illustration.}
%   \label{fig10}
%\end{figure}

\subsection{Constraint extraction}
Requirements + Metamodel $\rightarrow$ OCL constraints (Fig. \ref{fig12})

\subsection{Deployment configuration generation}

\begin{figure*}[t]
    \centering
    \includegraphics[width=0.5\textwidth]{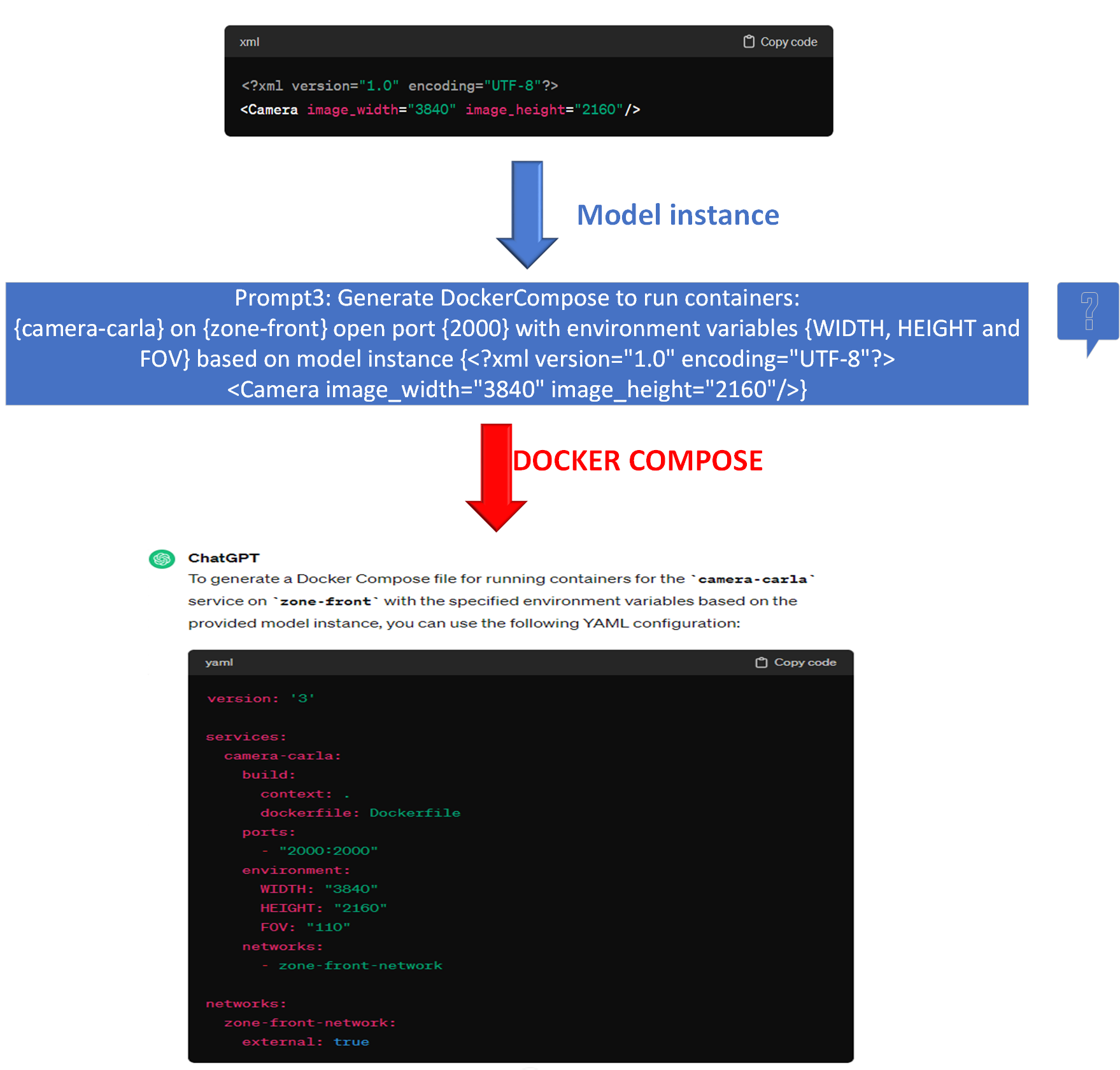}
    \caption{Deployment configuration generation experiment.}
   \label{fig11}
\end{figure*}

Model instance $\rightarrow$ Docker Compose (Fig. \ref{fig11})
%\begin{figure}[ht]
%    \includegraphics[width=\linewidth]{exp6.png}
%   \caption{Deployment configuration generation experiment.}
%   \label{fig11}
%\end{figure}
\subsection{Device interoperability}
\begin{figure*}[t]
    \centering
    \includegraphics[width=0.5\textwidth]{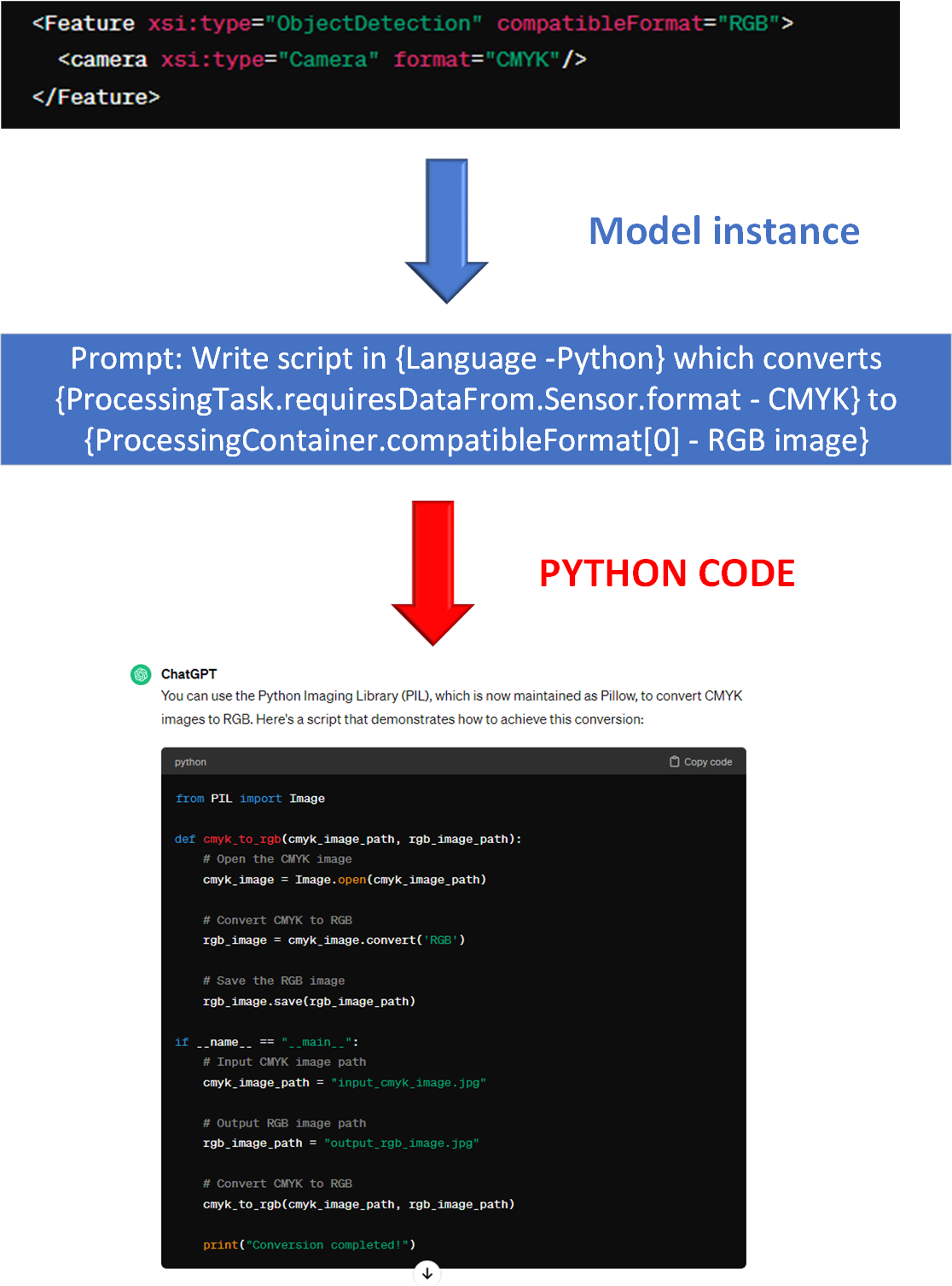}
    \caption{Data format interoperability experiment.}
   \label{fig12}
\end{figure*}

Model instance $\rightarrow$ Python adapter code  (Fig. \ref{fig12})
%\begin{figure}[ht]
%    \includegraphics[width=\linewidth]{exp7.png}
%    \caption{Data format interoperability experiment.}
%   \label{fig12}
%\end{figure}

\subsection{Requirements verification}
For purpose of formal design-time model instance verification, we make use of the so-called Object Constraint Language (OCL) declarative language. It represents an extension of UML notation, giving the ability to specify some system design aspects with more details, which are verification rules in our case. Using OCL notation, a set of logical rules based on reference architecture/ISO standard is defined which are checked against the user-provided model instance. An example of ISO standard which can be used for this purpose is ISO/TR 4804 \cite{ref48} (going to be replaced soon by ISO/CD TS 5083 \cite{ref49}). In our case, OCL rules are generated relying on LLM, starting from free-form text.
Let us assume that we start from reference hardware requirements for Conditional Automated Highway Drive: "This system represents an example sensor configuration with six cameras (at least 2.1MP), two LIDARs and five radars...".
In OCL notation, we would have the following rules for verification:
\begin{verbatim}
context Feature
inv CamerasCount:
self.cameras->size() = 6 
context Feature
inv LIDARCount:
self.lidars->size() = 2 
context Feature
inv RADARCount:
self.radars->size() = 5 
context LIDAR
inv LIDARChannels:
channels > 0
inv LIDARRange:
range > 0 
context RADAR
inv RADARRange:
range > 0 
context Camera
inv CameraResolution:
width > 0 and height > 0
inv MinimumResolution:
(width * height) >= 2.1 * 1000000  
-- Assuming resolution is measured
-- in megapixels (MP) 
\end{verbatim}

\section{Dataset}
For purpose of experimentation, we created a custom dataset of mock automotive industry requirements based on AVC Consortium document \cite{ref50}, which is publicly available. The reason for this decision comes from fact that most of the automotive vendors, OEMs and other industry partners might be concerned about their data becoming public, which could lead to potential legal issues.
As for now, two custom datasets are created: 1) reference requirements + Ecore metamodel -\textgreater OCL constraints 2) user requirements + Ecore metamodel -\textgreater Ecore model instance.

Table I depicts the layout of the dataset. We have the following inputs: 1) Requirement text -- Natural language, unstructured text covering some requirement defined within either the ISO standard or reference architecture used for vehicle system development (first dataset) or user-defined requirement (second dataset). Regarding the first dataset, the idea is to collect this kind of requirements from publicly available examples, reference documents or ISO standards for automotive industry. This way, we aim to enforce the adoption of good practices within the user-created system instances. On the other side, the requirements of the second type are collected from application form filled by user; 2) Feature -- to which feature the particular requirement is related to; 3) Scenario -- to which higher level functionality the particular feature belongs to (such as emergency break within highway pilot); 4) metamodel -- reference, human-created metamodel.

On the other side, we have two possible outputs: 1) OCL rules -- free-form text about the architectural and standardisation-related constraints expressed in form of OCL rules with respect to the input metamodel; 2) model instance -- the resulting annotation of free-form text with respect to the input metamodel.

\begin{table}[]
\caption{Layout (header) of the dataset: Inputs (Requirements text, Feature, Scenario, Metamodel) are marked white, outputs (OCL rule, Model instance) are marked gray.}
\begin{tabular}{|l|l|l|l|
>{\columncolor[HTML]{C0C0C0}}l |
>{\columncolor[HTML]{9B9B9B}}l |}
\hline
\begin{tabular}[c]{@{}l@{}}Requirement\\ text\end{tabular} & Feature & Scenario & Metamodel & OCL rule & \begin{tabular}[c]{@{}l@{}}Model \\ instance\end{tabular} \\ \hline
\end{tabular}
\end{table}

\section{Conclusion}
In this paper, we presented a prototype of model-driven framework leveraging LLMs, as an attempt to introduce automated software development to the automotive industry. Several use cases and examples adopting such development approach are demonstrated, with the main goal of generating an extensible and future-proof, centralized software vehicle system.

According to the initial findings, such approach seems promising, especially when it comes to reduction of time and effort needed for development of automotive software because of automated code generation. Additionally, automated verification of system models relying on Ecore model instance representation and OCL rules in design-time could potentially reduce the later efforts needed for run-time testing of the created systems. However, the presented workflow is a proof-of-concept, and it must be properly evaluated with large datasets of requirements used in the industry. Our plan is to incorporate more advanced techniques for processing of the ISO standards and reference architectures using LLMs, such as Retrieval Augmented Generation (RAG).

\clearpage

%\printglossaries

\printbibliography

@inproceedings{ref17:Sommer2013,
    author={Sommer, Stephan and Camek, Alexander and Becker, Klaus and Buckl, Christian and Zirkler, Andreas and Fiege, Ludger and Armbruster, Michael and Spiegelberg, Gernot and Knoll, Alois},
    booktitle={2013 IEEE International Electric Vehicle Conference (IEVC)}, 
    title={{RACE: A Centralized Platform Computer Based Architecture for Automotive Applications}}, 
    year={2013},
    volume={},
    number={},
    pages={1-6},
    doi={10.1109/IEVC.2013.6681152}
}

@online{ref30,
    title = {{Object Constraint Language}},
    year={2024},
    url = {https://www.omg.org/spec/OCL/2.4/About-OCL},
    urldate = {2024-02-19},
}

@Article{ref34:Azzoni2021,
    AUTHOR = {Al-Azzoni, Issam and Blank, Julian and Petrović, Nenad},
    TITLE = {{A Model-Driven Approach for Solving the Software Component Allocation Problem}},
    JOURNAL = {Algorithms},
    VOLUME = {14},
    YEAR = {2021},
    NUMBER = {12},
    ARTICLE-NUMBER = {354},
    ISSN = {1999-4893},
    DOI = {10.3390/a14120354}
}

@book{ref40:Beck2002,
    author = {Beck},
    title = {{Test Driven Development: By Example}},
    year = {2002},
    isbn = {0321146530},
    publisher = {Addison-Wesley Longman Publishing Co., Inc.},
    address = {USA}
}

@book{ref41:Palmer2001,
  title={A practical guide to feature-driven development},
  author={Palmer, Steve R and Felsing, Mac},
  year={2001},
  publisher={Pearson Education}
}

@BOOK{ref42:Incose2023,
  title     = "{INCOSE} Systems Engineering Handbook: A Guide for System Life Cycle Processes and Activities",
  editor    = "{INCOSE}",
  author    = {David D. Walden and Thomas M. Shortell and Garry J. Roedler and Bernardo A. Delicado and Odile Mornas and Yip Yew-Seng and David Endler},
  publisher = "John Wiley \& Sons Inc.",
  edition   =  5,
  month     =  jun,
  year      =  2023,
  address   = "New York",
  language  = "en"
}

@Book{ref43:Mitchell2002,
    author={Mitchell, Richard and McKim, Jim},
    title={Design by contract, by example},
    year={2002},
    publisher={Addison Wesley Boston, MA},
    address={Boston, MA},
    isbn={9780201634600},
    language={eng}
}

@INPROCEEDINGS{ref44:Pan2022,
  author={Pan, Fengjunjie and Lin, Jianjie and Rickert, Markus and Knoll, Alois},
  booktitle={2022 IEEE 25th International Conference on Intelligent Transportation Systems (ITSC)}, 
  title={{Resource Allocation in Software-Defined Vehicles: ILP Model Formulation and Solver Evaluation}}, 
  year={2022},
  volume={},
  number={},
  pages={2577-2584},
  doi={10.1109/ITSC55140.2022.9922526}}

@INPROCEEDINGS{ref45:Petrovic2023,
  author={Petrovic, Nenad and Al-Azzoni, Issam},
  booktitle= {6th International Conference on Applied Electromagnetics – PES 2023}, 
  title={{Automated Approach to Model-Driven Engineering Leveraging ChatGPT and Ecore}}, 
  year={2023},
  volume={},
  number={},
  pages={166-168} }

@INPROCEEDINGS{ref46:Petrovic_AlAzzoni_2023,
  author={Petrovic, Nenad and Al-Azzoni, Issam},
  booktitle= {ICSEng 2023}, 
  title={{Model-Driven Smart Contract Generation Leveraging ChatGPT}}, 
  year={2023},
  volume={},
  number={},
  pages={387-396},
  doi={10.1007/978-3-031-40579-2_37}}

@INPROCEEDINGS{ref47:Tas_2017,
  author={Ömer Şahin Taş and Stefan Hörmann and Bernd Schäufele and Florian Kuhnt},
  booktitle= {2017 IEEE 20th International Conference on Intelligent Transportation Systems (ITSC)}, 
  title={{Automated vehicle system architecture with performance assessment}}, 
  year={2017},
  volume={},
  number={},
  pages={1-8},
  doi={10.1109/ITSC.2017.8317862}}

@online{ref48,
    title = {{ISO/TR 4804:2020}},
    year={2024},
    url = {https://www.iso.org/standard/80363.html},
    urldate = {2024-03-17},
}

@online{ref49,
    title = {{ISO/CD TS 5083}},
    year={2024},
    url = {https://www.iso.org/standard/81920.html},
    urldate = {2024-03-17},
}

@online{ref50,
    title = {{AVC Consortium, Technical Reports \& More from AVCC }},
    year={2024},
    url = {https://avcc.org/documents/},
    urldate = {2024-03-17},
}

@techreport{ref51:Lebioda,
	author = {Krzysztof Lebioda and Viktor Vorobev and Nenad Petrovic and Fengjunjie Pan and Vahid Zolfaghari and Alois Knoll},
	title = {Towards Single-System Illusion in Software-Defined Vehicles - Automated, AI-Powered Workflow},
	year = {2024},
	 institution = {Technical University of Munich},

    url = {https://mediatum.ub.tum.de/doc/1737800/1737800.pdf}

}

@online{ref52,
    title = {{CARLA Simulator Sensors Reference}},
    year={2024},
    url = {https://carla.readthedocs.io/en/latest/ref_sensors/},
    urldate = {2024-03-17},
}

\end{document}